\documentclass[a4paper,10pt]{article}
\usepackage{graphicx}

\title{}
\author{}
\date{}
\begin{document}

\title{Picture of the low-dimensional structure in chaotic dripping faucets}% Force line breaks with \\

\author{Ken Kiyono, Tomoo Katsuyama, Takuya Masunaga,\\ Nobuko Fuchikami}
 
%\affiliation{Department of Physics, Tokyo Metropolitan University 1-1, Minami-Ohsawa, Hachoji, Tokyo 192-0397, Japan.}

\date{\today}% It is always \today, today,
\maketitle

\begin{abstract}
Chaotic dynamics of the dripping faucet was investigated both experimentally and theoretically. We measured continuous change in drop position and velocity using a high-speed camera. Continuous trajectories of a low-dimensional chaotic attractor were reconstructed from these data, which was not previously obtained but predicted in our fluid dynamic simulation. From the simulation, we further obtained an approximate potential function with only two variables, the drop mass and its position of the center of mass. The potential landscape helps one to understand intuitively how the dripping dynamics can exhibit low-dimensional chaos. 
\end{abstract}

It is well known that the beat of a dripping water faucet is not always regular and exhibits complex behaviour including chaos \cite{mar96}. Chaotic dripping was originally suggested by R\"ossler \cite{ros} and experimentally confirmed by Shaw and his collaborators \cite{sha85}. Since then, many experimental and theoretical studies have established the dripping faucet as a sort of paradigm for chaotic systems \cite{nun89,wu89,cah90,aus91,dre91,sar94,kat99,san95,inn98,ila99,ren99,lim97,fuc99,kiy99,pin00,amb00,kiy02}. Most previous studies have involved measuring the time interval $T_n$ between successive drips, because the dripping time is easily measured using a drop-counter apparatus \cite{sha85,nun89,wu89,cah90,aus91,dre91,sar94,kat99}. The time intervals are then plotted in pairs $(T_n, T_{n+1})$ for each $n$ to give a return map. Because the return maps typically appear low dimensional ($\sim$ one-dimensional), the behavior is often described by a simple dynamical model composed of a variable mass and a spring \cite{sha85,aus91,san95,inn98,ila99,ren99,lim97}. In this mass-spring model, a mass point, whose mass increases linearly with time at a given flow rate $Q$, oscillates with a fixed value of the spring constant $k$; and a part $\Delta m$ of the total mass $m$ is removed when the spring extension exceeds a threshold, which describes the breakup of a drop. Although the model exhibits chaotic return maps similar to those obtained experimentally, its empirical nature means that it does not provide a unified explanation for the complex behavior of the real dripping faucets, typically seen in bifurcation diagrams (plotting of $T_n$ vs.\ the control parameter $Q$). Thus, the connection between the low-dimensional dynamical system and a presumably infinite-dimensional fluid dynamical system remains elusive. 

Recently, we carried out fluid dynamic computations (FDC) based on a new algorithm involving Lagrangian description and succeeded in reproducing not only the time-dependent shapes of the drops, but also various characteristics of nonlinear dynamical systems, such as period-doubling, intermittency, hysteresis, observed in experiments of dripping faucets \cite{fuc99}. A detailed analysis of our FDC made it possible to improve the mass-spring model by using more realistic approximations including taking into account the mass dependence of the spring constant, and the necking process \cite{kiy99}. This improved mass-spring model, described only by three variables: mass $m$, its position $z_{\rm G}$ and the velocity $\dot{z}_{\rm G}$, closely described the dynamical behavior exhibited by the FDC. In particular, qualitatively good agreement among bifurcation diagrams (especially the repeating structure which reflects the oscillation frequency of each drop), obtained by (i) the experiment \cite{kat99}, (ii) the FDC \cite{fuc99} and (iii) the improved mass-spring model \cite{kiy99}, over a wide range of the control parameter $Q$ provided a theoretical model of the basic low-dimensional structure inherent in the system, which is presented here.

\begin{figure}
\begin{center}
\includegraphics[width=.9\linewidth]{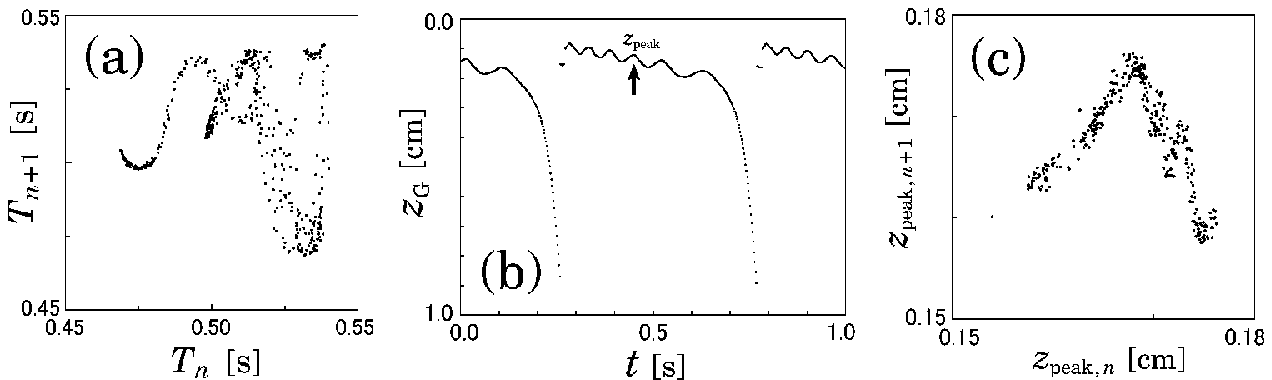}% Here is how to import EPS art
\caption{\label{fig:1}Experiments on the dripping nozzle (7 mm inner diameter and 10 mm outer diameter). The flow rate $Q = 0.24$ g/s ($\sim 2$ drips/s on average). (a) Return map of dripping time intervals, where the time intervals in pairs $(T_n, T_{n+1})$ for each $n$ are plotted. The time series $\{ T_n \}$  was measured using a drop-counter apparatus wherein a laser beam directed at a detector is interrupted whenever a drop crosses the beam. (b) Oscillation of the position $z_{\rm G}$ of the center of mass of the drop under the nozzle. $z_{\rm G}$ was estimated from the shape of the drop using the digitised image recorded by a high-speed video camera every 1/500 s. (c) Return map of the fourth peak $z_{{\rm peak}, n}$ of the oscillating $z_{\rm G}$, where $z_{{\rm peak}, n}$ is indicated by an arrow in (b). }
\end{center}
\end{figure}

An experimental chaotic attractor is shown in Fig.\ \ref{fig:1}. In addition to the discrete-time variable $T_n$, the mass $m(t)$ of the liquid suspended by the nozzle, the position of the center of mass $z_{\rm G} (t)$ and velocity $\dot{z}_{\rm G}$ were also measured. As seen in Fig.\ \ref{fig:1}(b), $z_{\rm G}(t)$ oscillates because the surface tension works as a restoring force. Similar oscillation was observed in both our FDC and improved mass-spring model \cite{kiy99}. Note that the frequency of the oscillation during each time interval $T_n$ is determined by the flow rate. At $Q = 0.24$ g/s, $z_{\rm G}$ oscillates $6 \pm 1/2$ times before the breaking up and the phase of the oscillation in the plane $(z_{\rm G}, \dot{z}_{\rm G})$ fluctuates chaotically at breakup. A return map of $\{z_{{\rm peak}, n} \}$ for every fourth peak of $z_{\rm G}(t)$ is presented in Fig.\ \ref{fig:1}(c). The return map of $T_n$ (Fig.\ \ref{fig:1}(a)) looks like an entangled string and thus the map function $T_{n+1} = G (T_n)$ cannot be single-valued. In contrast, the map function $z_{{\rm peak}, n+1} = F(z_{{\rm peak}, n})$ is approximately single-valued, although broadened by 'noise'. Low-dimensional multi-valued map functions of $T_n$ have often been obtained in both experiments \cite{sha85,nun89,wu89,kat99}, and numerical simulations \cite{fuc99,kiy99}. For the improved mass-spring model \cite{kiy99}, the map function $m_{{\rm r},n+1} = H(m_{{\rm r},n})$ ($m_{{\rm r},n}$ being the remnant mass just after the $n$-th breakup), as well as the map function $F(z_{{\rm peak}, n})$ is single-valued, while $G(T_n)$ is generally multi-valued. This is because the mapping of the time interval $T_n$ is not topologically conjugate to that of the dynamical variable $m_{{\rm r},n}$. The FDC results also showed that $H(m_{{\rm r},n})$ is approximately single-valued even when $G(T_n)$ is multi-valued. Thus, $T_n$ does not directly specify the state of the system despite its easy measurement. 

\begin{figure}
\begin{center}
\includegraphics[width=.5\linewidth]{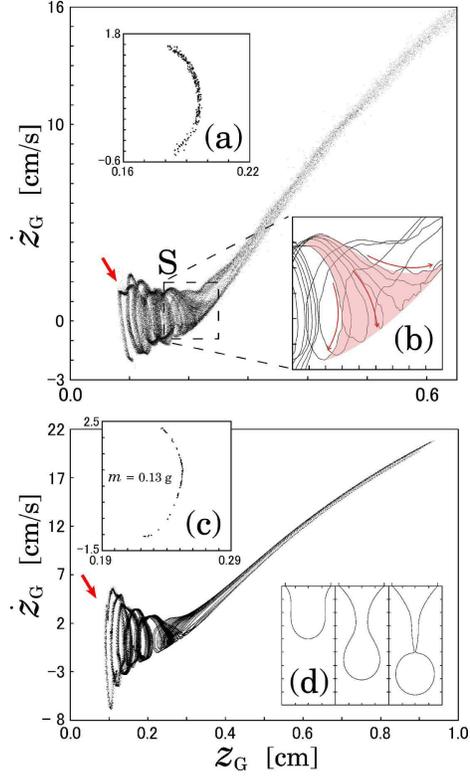}% Here is how to import EPS art
\caption{\label{fig:2}The top panel is the experimentally observed strange attractor under the same experimental conditions as in Fig.\ \ref{fig:1}, where the position $z_{\rm G}$ and the velocity $\dot{z}_{\rm G}$ of the drop under the nozzle are plotted at every 1/500 s. (a) Poincar\'e cross section of the attractor (plotting of $\dot{z}_{\rm G}$ vs.\ $z_{\rm G}$) at $t = 0.3$ s measured from each breakup moment; (b) Enlarged view of the region S in which the attractor undergoes stretching and folding. The bottom panel is numerical simulations, where water drops falling from a nozzle (7 mm diameter) are simulated at the flow rate $Q = 0.32$ g/s. We have used the same parameter values as in Ref. \cite{kiy99}. (c) Poincar\'e section of the attractor ($\dot{z}_{\rm G}$ vs.\ $z_{\rm G}$) at $m = 0.13$ g; (d) Drop deformation. }
\end{center}
\end{figure}

Measurement of the continuous-time variables $ \{ z_{\rm G}(t), \dot{z}_{\rm G}(t), m(t) \} $ made it possible to visualize the trajectory of the chaotic attractor in a continuous state space for the first time. 
The projection of the attractor in the plane $(z_{\rm G}, \dot{z}_{\rm G})$ corresponding to the chaotic motion in Fig.\ \ref{fig:1} is shown in Fig.\ \ref{fig:2}(top). 
On average, $z_{\rm G}(t)$ oscillated six times during each time interval $T_n$ (see Figs.\ \ref{fig:1}(b) and \ref{fig:2}(top)). The cross section of the attractor in Fig.\ \ref{fig:1}(a) looks nearly one-dimensional and explicitly shows that the drop formation is well characterized by a few state variables. Figure \ref{fig:2}(b) is a blow-up of the region S in which trajectories starting with slightly different remnant masses separate rapidly from each other. The transition from the oscillating process to the 'necking' process (i.e., stretching of the attractor) occurs in this region S \cite{kiy99}. In Fig.\ \ref{fig:2}(bottom) the FDC results are presented including profiles of the evolving drops (Fig.\ \ref{fig:2}(d)), which show qualitatively good agreement with the experimental observations. The cross section of the theoretical attractor (Fig.\ \ref{fig:2}(c)) also exhibits the low-dimensionality. An attractor in the space $(z_{\rm G}, \dot{z}_{\rm G}, m)$ with a similar spiral structure has already been obtained using the improved mass-spring model \cite{kiy99}. These results strongly support the use of the three state variables:  $z_{\rm G}$, $\dot{z}_{\rm G}$  and $m$ to describe the drop motion. 

\begin{figure}
\begin{center}
\includegraphics[width=.9\linewidth]{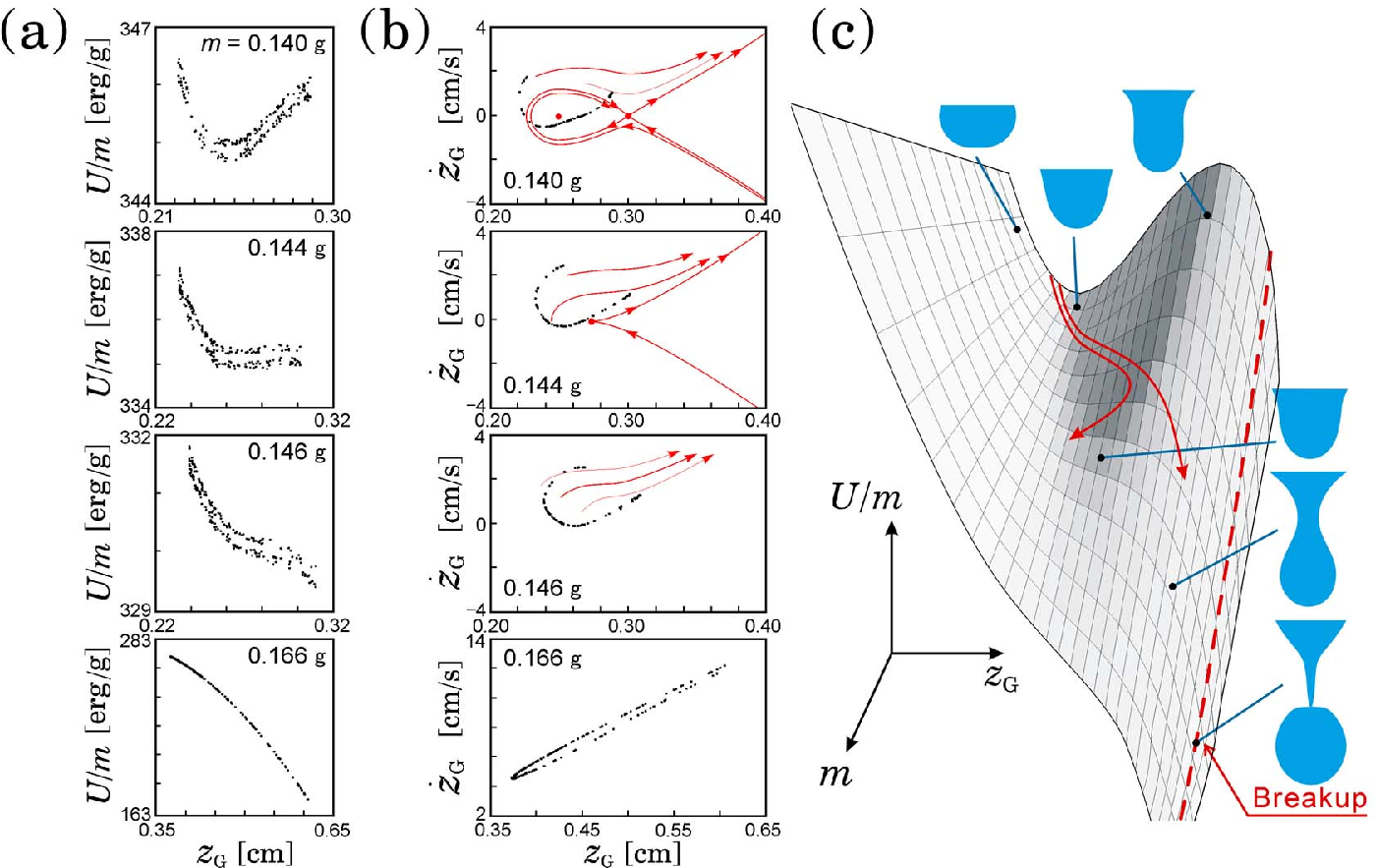}% Here is how to import EPS art
\caption{\label{fig:3}(a) Fluid dynamic simulation (7 mm diameter, the flow rate $Q = 0.32$ g/s). Cross section of $U/m$ for the attractor is plotted vs.\ $z_{\rm G}$ on a constant mass surface, where $U/m$ is the total potential (= surface tension plus gravitation) energy per unit mass as a function of $m$ (drop mass in units of gram in each panel) and $z_{\rm G}$ (position of the center of mass). The critical mass $m_{\rm crit} = 0.144$ g. (b) Black dots represent the Poincar\'e section of the attractor, plotting of $\dot{z}_{\rm G}$ vs.\ $z_{\rm G}$, on the same constant mass surface as in (a) (obtained from the same simulation). The red solid curves indicate the flow in the state space, which were drawn by referring to the results obtained from another simulation with fixed mass (i.e., $Q = 0$). The red points in the top two panels are fixed points for the $Q = 0$ system. (c) Schematic view of a typical potential surface, obtained by approximating $U/m$ in (a) by a single valued function of $z_{\rm G}$ and $m$. The drop shapes (light blue) in various state points $(z_{\rm G}, m)$ on the potential surface are indicated. The red lines represent the trajectories starting with slightly different masses; one oscillates in the potential valley, while the other gets over the ridge because of a slight difference of the velocity $\dot{z}_{\rm G}$. The breakup occurs on the red dashed line. }
\end{center}
\end{figure}

The motion of the drop is subjected to gravitational force and surface tension. Since the surface energy depends on the shape of the drop, the total potential energy $U$ (i.e., gravitational plus surface) should be a function of the many degrees of freedom of the liquid. Our FDC show, however, that approximating $U$ as a function of two variables, $m$ and $z_{\rm G}$, yields a conceptually clear picture for the basic low-dimensional structure of the system. Cross sections of $U/m$, the potential energy per unit mass, at several fixed values of $m$ for the chaotic attractor shown in Fig.\ \ref{fig:2}(bottom) are presented in Fig.\ \ref{fig:3}(a). Poincar\'e cross sections at the same values of $m$ as Fig.\ \ref{fig:3}(a) are given in Fig.\ \ref{fig:3}(b). Although the cross sections in Fig.\ \ref{fig:3}(a) look two-folded, if they are approximated as single-valued functions, then a sheet of the potential surface $U(m, z_{\rm G})/m$ is obtained as shown in Fig.\ \ref{fig:3}(c). The surface is characterized by a U-shaped valley and a ridge which converge as $m$ increases and have totally merged when $m = m_{\rm crit}$, the maximum mass of the static stable shape. We have also carried out a numerical experiment based on FDC with $Q = 0$ by fixing the value of $m$. Various artificial shapes of the drops were used as initial conditions, and it was found that the point $(m, z_{\rm G})$ rapidly converged onto a well-defined potential surface $U(m, z_{\rm G})/m$ similar to that shown in Fig.\ \ref{fig:3}(c) \cite{note01}. 

The potential surface for the closed system (i.e., the pendant drop for $Q = 0$) is useful for understanding the dynamics of the open system ($Q > 0$). The minimum and a maximum of $U(m, z_{\rm G})/m$ for the $Q = 0$ system correspond to a stable fixed point (static equilibrium shape) and a saddle point (unstable equilibrium shape, see Fig.\ \ref{fig:3}(c)), respectively. These are indicated by red points in the top two panels of Fig.\ \ref{fig:3}(b). When $m = m_{\rm crit}$ ($= 0.144$ g), the stable fixed point disappears, and the necking process begins leading to the pinching off of the drop. When $m < m_{\rm crit}$, the trajectories of the drops oscillate in the valley of $U(m, z_{\rm G})/m$, which corresponds to the spiral structure of the attractor in Fig.\ 2. The flow of the $Q = 0$ system as depicted in Fig.\ \ref{fig:3}(b), also suggests that the state point of the drop causes the spiral orbits in the phase space $(z_{\rm G}, \dot{z}_{\rm G}, m)$. The chaotic dynamics, especially the stretching and folding of the attractor, can be interpreted as follows. The stable fixed point approaches the saddle point as $m$ increases. The state point thus visits the vicinity of the saddle point after the spiral motion, which causes the instability of the orbit, or equivalently, the stretching of the attractor (as in Fig.\ \ref{fig:2}(b)). The Poincar\'e section of the chaotic attractor (Fig.\ \ref{fig:3}(b)), which rotates around the stable fixed point with increasing $m$, is gradually crushed and asymptotes to a single line through the necking process. 

The above analysis leads to the motion of the dripping faucet being described in terms of the approximate potential function $U$ as
\begin{equation}
\ddot{z}_{\rm G} = - \frac{\partial U(m, z_{\rm G})/m}{\partial z_{\rm G}} - \frac{(\gamma + Q)}{m} \dot{z}_{\rm G} \quad,\ \dot{m} = Q \label{eq:m},
\end{equation}
where $\gamma$ is the damping constant. This equation describes the motion of a particle with unit mass in 2-dimensional space $(m, z_{\rm G})$, where the particle is forced to have a constant velocity $\dot{m} = Q$ in the direction of the $m$-axis. Our improved mass-spring model corresponds to a local quadratic approximation for the potential near the bottom $z_{\rm G} = z_0$ of the valley in which the $m$-dependence of the spring constant $k(m) \equiv \partial ^2 U(m, z_0) / \partial z_0 ^2 $ has been included \cite{kiy99}. The present description is thus a natural extension of the improved mass-spring model. As we have seen, however, the picture of the global landscape of the potential is much clearer than the improved mass-spring model for intuitively understanding the dynamics of the system. If the flow rate $Q$ is small enough so that the initial oscillation after the breakup is damped, the particle (equivalently the state point of the drop) goes along the bottom of the valley as $m$ increases. At $m = m_{\rm crit}$, where the valley merges into the ridge, the particle's motion loses its stability (start of the necking process) and rapidly approaches the breakup point (on the broken red line in Fig.\ \ref{fig:3}(c)). Since the instability always begins at the same point $m = m_{\rm crit}$ in this case, the size of each drop is uniform. If, on the other hand, the oscillation of the drop affects the start of necking, the dripping motion exhibits a variety of flow rate-dependent periodic and chaotic patterns. Note that the particle generally gets over the ridge before $m$ reaches $m_{\rm crit}$ when it is oscillating. At a flow rate resulting in a chaotic motion, two trajectories starting with slightly different $m$ values (different remnant masses) are initially close to each other. As $m$ increases, however, one trajectory may get over the ridge at a certain $m$ value, while the other remains in the valley owing to a small difference in $\dot{z}_{\rm G}$  (two red lines in Fig.\ \ref{fig:3}(c)). 

Similar potential landscapes are expected for other subjects involving drop formation like ink-jet printing \cite{shi87} and atomization \cite{ama98}. The potential surface describing the motion in the attractor was confirmed by FDC to have essentially the same topology even for high viscosity liquids and for microfluidic systems (drop size $\sim$ pl). Hence, the present picture of a particle moving along a low-dimensional potential surface is relevant to modeling drop formation in general. 

We are grateful to Professor W. S. Price and Professor A. L. Salas-Brito for careful reading of the manuscript.

\newpage

\begin{thebibliography}{99}
\bibitem{note01}
Besides the up and down-oscillation of the center of mass of the liquid, there are other oscillation modes which induce drop deformation. These extra modes result in the fine structures of the attractor. For example, the map function in Fig.\ 1 (c) shows two or three peaks apart from the main one. Close observation of the experimental and FDC results revealed that these fine structures correspond to the deformation modes of the liquid. A deformation mode can be seen in Fig.\ 2 as a sharp bend of the trajectory (red arrows). This extra mode, which has not been previously observed, was found to be a collision of the pendant drop with the nozzle, triggered by the shock of breakup. The mode is unstable and rapidly decays and thus only affects the initial oscillation just after the break. Because of the excitation of the extra mode, the state point $(m, z_{\rm G})$ separates from the well-defined surface $U(m, z_{\rm G})$ at the initial stage, and rapidly returns ($\sim $ ms) to the surface. This process stretches and folds the chaotic orbit, which makes the potential surface two-folded as in Fig.\ 3 (a).
\end{thebibliography}
\end{document}